# Rebellion on Sugarscape: Case Studies for Greed and Grievance Theory of Civil Conflicts using Agent-Based Models

Rong Pan

Arizona State University, Tempe, Arizona, United States
rong.pan@asu.edu

**Abstract.** Public policy making has direct and indirect impacts on social behaviors. However, using system dynamics model alone to assess these impacts fails to consider the interactions among social elements, thus may produce doubtful conclusions. In this study, we examine the political science theory of greed and grievance in modeling civil conflicts. An agent-based model is built based on an existing rebellion model in Netlogo. The modifications and improvements in our model are elaborated. Several case studies are used to demonstrate the use of our model for investigating emergent phenomena and implications of governmental policies.

## 1 Introduction

In this research we are interested in modeling civil conflicts using the agent-based computational approach, so as to study the behavior of the populace in a specific national environment and to investigate the impact of governmental policies. Based on the greed and grievance theory, a person rebels because 1) he has enormous grievance towards the central government and 2) he sees the opportunities (e.g., economical benefits, political advantages, etc.) of becoming a rebel. Therefore, we may say that the occurrence of civil conflict is the result of the interaction of populace within its environment, including the reactions of individuals to their life conditions and their interactions with neighbors, with the general national status (weak or strong state) and the authoritarian power of government, etc. Agent-based models (ABMs) are the natural choice of modeling such complex systems.

The NOEM model, developed by the Air Force Rome Laboratory, is a large-scale computer simulation model built on system dynamics principles for simulating nation-state operational environment. Through it we may study the causal relationship between exogenous factors and nation-state's social and economical status, as well as the impact of various policy options. We postulate that the NOEM will provide a valid nation-state environment for ABM simulation. For example, the indicators for the economic conditions of a nation, such as GDP, GINI, can be imported from the NOEM and they are used in ABMs to create a nation's economy map. The agents in ABMs will possess certain properties that link them to their living environment. They also follow certain behavior rules and interact with the environment. The NOEM is a system dynamics (SD) model, where macroscopic difference equations are used to calculate the change of nation-state variables. The ABM technique takes a different approach. It models at the microscopic level, modeling individual agent's (it can be a person) behavior, and how it interacts with its neighbors and its environment.

## 2 Theory of greed and grievance in civil conflicts

In the seminal paper by Collier and Hoeffler (2002), the outbreak of civil war was explained from two possible views: severe grievance and atypical opportunity (greed) for building a rebel organization. The grievance model is in line with the motivation cause of conflict explained in the political science literature. Grievance is fueled by the polarization of a society, which is caused by poor governance. The opportunity model explains the outbreak of civil war from an economic feasibility point of view. An interesting hypothesis proposed in Collier, Hoeffler and Rohner (2008) states that where civil war is feasible it will occur without reference to motivation. This "feasibility hypothesis" is substantiated by analyzing past civil war data and finding that the variables, which are close proxies for feasibility, have powerful consequences to the risk of a civil war.

Collier and Hoeffler's theory has stirred a heated debate on the true causes of conflicts among political scientists, sociologists and economists. Although most empirical studies conducted at the World Bank support the greed model (see, Ganesan and Vines, 2004, Bodea and Elbadawi, 2007), many researchers rebutted the notion that the onset of conflict is purely determined by economic reasons or the motivation of rebellion is driven by seeking financial opportunity only. Some argued that the reductionist categories of greed and grievance not only obfuscate other social, ethnic and religious factors, but also mislead the intervention policy of international community (Kuran, 1989, Gurr and Moore, 1997, Gurr, 2000, Ballentine and Nitzschke, 2003, Regan and Norton, 2003). This is specifically demonstrated by studying conflicts in some African countries (Agbonifo, 2004, and Abdullah, 2006). It seems that this debate will continue. In our study, we do not intend to validate this theory, but simply apply it on an artificial society and to observe the interactions of greed and grievance factors and the social consequence of governmental policy.

## 3 Agent-based models

An agent-based model (ABM) is a simulation environment where agents (e.g., simulated people) are defined to possess certain properties and behaviors, and they can interact with each other and with their surrounding environment. This type of model is different from other mathematical models, such as system dynamics, in several ways. First, the ABM is built in a bottom-up fashion. It defines the microscopic details of basic elements in a society and lets the society grow (or evolve) through agent-to-agent and agent-to-environment interactions. Second, the use of ABMs is typically to provide the macroscopic picture of how a society grows and look for emergent phenomena. This is achieved by examining the statistics of collections of agents over the simulation time. The connections between micro-level details and macro-level phenomena can be used to explain cause-and-effects for the artificial society under study.

Naturally, ABMs have been applied in many social science studies. See Epstein (1990) for supporting arguments of this computational approach to social science. Epstein and Axtell (1996) gave a full-fledged description of how to build an artificial society and examine its emergent phenomena on a sugarscape (an imagined land which provides sugars for the agents living on it). For the civil conflict study, the most famous model is the rebellion model, which implements the greed and grievance theory on modeling agent's behavior. It has been used to demonstrate certain conflict phenomena found in the real world and the effects of changing governmental policy (Epstein et al., 2001, Epstein, 2002, and Cederman and Girardin, 2007). More recently, a group in Netherlands used ABMs to simulate the war game scenario of foreign-force evasion on an island undergoing a civil war (van Lieburg et al. 2009, Borgers et al. 2009).

In our study, we built an artificial land based on the sugarscape model and let rebels grow on this land and interact with the environment; thus, we call it "rebellion on sugarscape". The sugarscape can be viewed as a resource map which is geospatially bounded. The behavior of an agent depends on the agent's living environment (whether there are enough resources to support its living and whether there are rebels/cops in its neighborhood) and the policy set imposed on this artificial land. Our intention is to use this example to demonstrate the feasibility of the marriage of civil conflict theory and a resource map, which has not been done in the original rebellion model. In addition, we studied the behavior of rebellion and the consequences of policy setting when the resource is pre-defined.

### 3.1 Rebellion on sugarscape model

Although the rebellion model has been successfully used to explain some phenomena of civil conflicts, it lacks several aspects that we are interested in. First, the grievance level of each person is randomly given. In modeling a nation, we would like to link the grievance level to the person's economic, social, and political status, or in other words, its quality of life. Second, the number of cops is fixed. In reality, a government needs to spend enormous wealth on maintaining its military, police and other civil governing means. Third, in the current model, all agents can freely, randomly move. In reality, there may be many restrictions on a person's movement for economic and political reasons.

To address these problems, we created a refined model, called "rebellion on sugarscape". The sugarscape is a layer of resource overlaid on the artificial land. Sugars (or resources) constantly grow on this land and can be collected by each agent, and by the government through taxation, to become personal wealth or the government's wealth. However, the distributions of sugars are uneven. There are areas the sugar level is highest (4) and areas there is no sugar (0). We may interpret the sugar-rich region as the resource-rich city area and the sugar-poor region as the resource-depleted rural area, or the sugar-rich region as the region with oil fields and the sugar-poor region as the region without oil. Initially, agents are randomly scattered on the land. An agent's grievance will be linked to the location it resides. If it is in a sugar-rich region, its perceived hardship will be low, and vice versa. The general population can be defined either mobile or non-mobile. If it is mobile, a person will move to a place having at least a sugar level of 1 within its vision. This is based on the assumption that people want to seek a better life for themselves. Every person will collect wealth from the place it resides. The wealth it can collect is the same as the sugar level at its location. Part of the wealth will be taxed, and the government will use the tax revenue to either maintain the existing cops, or to create new cops, or to assist the people who are in poverty. In our model, a cop needs at least a value of 5-sugar level to survive. To create a new cop, the government needs to spend 10 sugars. The poverty line is set at 1-sugar level. If a person has less than 1 sugar, it may be assisted by the government. The assisted amount is the total sugar that the government allocated to the assistance divided by the total number of people in poverty. Our model is depicted in Figure 1.

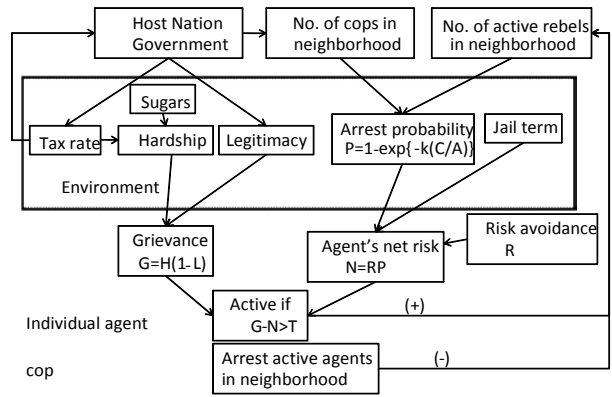

**Fig. 1.** The rebellion on sugarscape model

## 4 Case studies

In this section, we present several case studies of the rebellion on sugarscape model. Our purposes are two-folded: 1) to validate the model against basic, intuitive scenarios; 2) to look for "surprises" that will lead to a better understanding of the model and of the consequences of policy setting.

### 4.1 Case study #1: A baseline case

First we test a baseline case, where the number of agents (general population) is 1750 and the initial number of cops is 100. We prohibit the movement of agents (presenting a non-mobile society), and set both the wealth distributed to the poor and the wealth for creating new cops to be 0. Therefore, the government will use its wealth to maintain the initial cop size throughout the simulation. The tax rate is set at 30%.

This baseline case is a simple recast of the rebellion model using a layer that defines the resources and links the people's grievance to their wealth. Therefore, we should expect a similar behavior of rebels as in the previous rebellion simulation, except that the rebels will be most likely be in poor regions. We ran the simulation over 200 cycles. We observed that rebellions always happen in the places with less resource and the outburst of mass rebellion happens from time to time, and then it is countered by the cops. These periodic rebellion outbursts are similar to what has been described in the original rebellion model.

**4.2 Case study #2: Building cops vs. assisting the poor**

To suppress rebels, there are two basic strategies that the government can use – increasing the number of cops and providing financial assistance to the poor. In the following simulations, we tried both strategies and their combinations.

First, we set the Wealth distribution=0 and Wealth to create cops=0.5. This implies that the government will send half of its wealth to maintain the existing cop force and half of it to increase cops, and no assistance to the poor. As some cops will die due to lack of sugars, the number of cops will eventually be stabilized at 158. There are occasionally one or two rebels, but they will be quickly detained by the cops surrounding them. Over time, the number of jailed individuals varies around 18 to 20.

Next, we set Wealth to create cops=1.0, so all of the wealth the government collects will be spent on building new cops. Surprisingly, in the simulation, the cop size drops to 90 and a constant mass of rebels is created in the poor regions. We observe that a poor person is likely to choose cycling actions of being dormant, then actively rebel, and then be jailed. It is because they receive no help from the government and they cannot improve their lives by moving to rich regions, and in addition, there are not enough cops to suppress rebellions. The drop of cop force is surprising, given that the government wants to create more cops. But, in fact, most of the existing cops will die due to lack of sugar supply from the government, the new cops built by the government cannot keep up the pace of the dying cops. Increasing the number of new cops is more expensive, and the new cops are located randomly (less experience), less effective countering the rebels in the poor regions. Also, with more people put into jail, less tax revenue will be collected by the government, thus less funds for the cops. Moreover, from this simulation we observe that the periodic rebellion outbursts have been replaced by constant rebellions, which indicates the overall lack of a policing force.

However, when we set Wealth distribution=0.5 (half of the wealth will be redistributed to the people under the poverty line) and Wealth to create cops=0, we observe that there is still enough wealth to support 100 cops. In fact, there are very few incidences of rebellion, and even if there is an active rebel, these will be quickly detained by the cops. Over the course of the simulation, the number of jailed individuals varies from 0 to 5. This is a peaceful society. The poor will be assisted by the government, so their grievance levels cannot be too high. Also, the government will maintain a sizable force of cops.

### 4.3 Other cases

**Table 1.** Summary of the cases being studied and their emerging phenomena

| Case | Parameter | Emerging phenomenon |
|---|---|---|
| Baseline case | As in Section 4.1 | Periodic outburst of rebellion as in the rebellion model |
| Increasing the number of cops | wealth-distribution=0, wealth-to-create-cop=0.5 | Number of cops increases, but still have rebels from time to time |
|  | wealth-distribution=0, wealth-to-create-cop=1 | Number of cops decreases, have a large number of rebels |
| Assisting the poor | wealth-distribution=0.5, wealth-to-create-cop=0 | Very few rebels, the number of cops is maintained at the initial level |
| Freedom to move | movement?=on, wealth-distribution=0, wealth-to-create-cop=0 | People move to rich regions, cops move to poor regions, rebellion happens at the boundaries of rich and poor regions |
|  | movement?=on, wealth-distribution=0.5, wealth-to-create-cop=0.2 | No rebel |
| Less population | number of agents=250, initial number of cops=15, movement?=on, wealth-distribution=0 | The percentage of population who rebels increases, the frequency of rebellion increases |
|  | number of agents=250, initial number of cops=15, movement?=on, wealth-distribution=0.5 | No rebel |
|  | number of agents=250, initial number of cops=15, movement?=on, wealth-distribution=0, wealth-to-create-cop=0.5 | The cop number increases to 22, but still have rebels in less wealthy regions |
| High tax | tax=0.5 | Massive rebels in less wealthy regions even when the government redistributes some of its wealth to the poor |
| Less education | vision=3 patches | Some people will be isolated in poor regions, rebels happen at these poor regions, government assistance will help reduce rebels |
| Less government legitimacy | government-legitimacy=0.35 | More rebels than quiet population, cop number is reduced, the government collapses. The legitimacy level of 0.75 |

is the boundary value for maintaining an effective government.

## 5. Some issues for future research

In this simple artificial society we simulated, an agent's grievance is determined by its perceived hardship and its perceived government legitimacy only. Although we are able to link the agent's perceived hardship to its wealth (sugar level), the model is far away from a real world where people have various motivations for becoming rebels. Regarding to the motivations of violent non-state actors (VNSA), Bartolomei et al. (2004) postulated a framework of summarizing grievance from the perspective of human needs. On the national level, we may identify a group of variables that are important to the health of the nation, but to use ABMs we have to associate these macroscopic measurements with microscopic perceptions of individual agents. Little research has been done on this front so far, but this is a critical link between SD model and ABM.

Finally the details of how to utilize the NOEM as a nation-state environment for ABMs still need to be researched. The overall architecture that we envisioned is that from the NOEM we can create multiple layers of maps, corresponding to the nationals economic, social and political variables, and use them to define the properties and behavior rules of agents living on the artificial land. However, to materialize this concept, in-depth investigations of the NOEM's capabilities and its potential use for ABMs are required.

## Acknowledgement:

The author thanks the AFRL Summer Faculty Fellowship Program for its support of this research, and thanks Dr. John Salerno for his advising during the author's stay at the Air Force Rome Laboratory in the summers of 2009 and 2010.